\documentclass[twocolumn,secnumarabic,amssymb,nobibnotes,aps,prd,showpacs]{revtex4-1}

\usepackage{graphicx}
\newcommand{\be}{\begin{equation}} 
\newcommand{\ee}{\end{equation}} 
\newcommand{\bea}{\begin{eqnarray}} 
\newcommand{\eea}{\end{eqnarray}} 
\newcommand{\eml}{\end{mathletters}} 
\newcommand{\nn}{\nonumber\\} 
\newcommand{\oh}{\frac{1}{2}}

\newcommand{\la}{\langle} 
\newcommand{\ra}{\rangle}

\newcommand{\what}{\widehat}

\begin{document} 

\title{Cluster mean field description of alpha emission}

\author{A. Dumitrescu $^{1,2,3}$ and D.S. Delion $^{1,2,4}$}
\affiliation{
$^1$ "Horia Hulubei" National Institute of Physics and Nuclear Engineering, \\
30 Reactorului, POB MG-6, RO-077125, Bucharest-M\u agurele, Rom\^ania \\
$^2$ Academy of Romanian Scientists, 3 Ilfov RO-050044,
Bucharest, Rom\^ania \\
$^3$ Department of Physics, University of Bucharest, 405 Atomistilor, POB MG-11, RO-077125, Bucharest-M\u agurele, Rom\^{a}nia \\
$^4$ Bioterra University, 81 G\^arlei RO-013724, Bucharest, Rom\^ania}
\date{\today}

\begin{abstract}
We show that the Hartree--Fock--Bogoliubov (HFB) method is able to describe
experimental values of alpha decay widths by including a residual 
nucleon--nucleon Surface Gaussian Interaction (SGI) within the standard 
procedure used to calculate the nuclear mean field. 
We call this method the Cluster HFB (CHFB) approach.
In this way we correct the deficient asymptotic behaviour of the corresponding 
single--particle (sp) wave functions generated by the standard mean field. 
The corrected mean field becomes a sum between the standard mean Woods--Saxon--like
field and a cluster Gaussian component centered at the same radius as the SGI.
Thus, we give a confirmation of the mean field plus cluster potential structure, 
which was assumed in our previous work on alpha-decay widths.
Systematic calculations evidence the linear correlation between the SGI strength
and fragmentation potential, allowing for reliable predictions concerning
the half lives of superheavy emitters.
\end{abstract}


\maketitle

\section{Introduction} 
\label{sec:intro} 
\setcounter{equation}{0}
\renewcommand{\theequation}{1.\arabic{equation}} 

From the very first theories of $\alpha$--emission published by Gamow \cite{Gam28} and
independently by Condon and Gurney \cite{Con28}, almost a century passed until 
$\alpha$--particles were experimentally observed on the surface of nuclei \cite{Tan21}.
However, describing the formation of $\alpha$--particles on the surface of an atomic nucleus from
two protons and two neutrons still remains a considerable theoretical challenge within the microscopic 
theory of $\alpha$--decay from heavy nuclei.
This radioactive process is fundamental in explaining the dynamics of various exotic physical systems, 
like supearheavy and highly unstable nuclei \cite{Gam05}. 
The estimations of absolute $\alpha$-decay widths, where only one shell
model configuration was considered, were smaller than the 
experimental data by several orders of magnitude \cite{Man60,San62}. 
The typical example is the decay process $^{212}$Po $\rightarrow ^{208}$Pb + $\alpha$,
where two proton and two neutron orbitals were considered above the doubly magic $^{208}$Pb. 
It was soon realized that by increasing the number of single--particle configurations the value of
the decay width substantially increases \cite{Sol62,Man64}. But even if a 
very large number of shells was included in order to simulate the continuum part of the spectrum,
the absolute decay width still deviated by more than one order of magnitude
\cite{Fli76,Ton79}. The reason why the absolute decay width increases
with the number of configurations is due to the clustering
of the nucleons forming $\alpha$-particles, implying the inclusion of high lying
configurations in the formation process \cite{Jan83}.
Even so, the calculated absolute decay widths still differed frome experimental
observations by at least one order of magnitude \cite{Del92,Del00,Len93,Bet12}.

The phenomenological model used to solve this problem consists in representing the emission process 
through a cluster moving in an attractive pocket-like potential located at the nuclear surface.
Under the assumptions of the R--matrix theory \cite{Lan58},
this model predicts an analytic linear dependence between the logarithm of the
reduced width and the fragmentation potential, defined by the difference between the Coulomb barrier
and Q-value \cite{Del09}. It remains valid for many strong emission processes, 
including proton radioactivity and heavy cluster decay \cite{Dum22}.
This indicates that the representation we are seeking must be provided by an
attractive potential like that ``pocket'' potential in addition to the standard
Woods--Saxon plus spin--orbit mean field. Furthermore, it is interesting to point
out the existence of an alternative description of clustering phenomena employing
the nonlinear Schr\"{o}dinger equation and solitions on quantum droplets \cite{Car21}. 

The idea of extending the description of nuclear interactions beyond the mean field is not new, but in this work we show that the proposed potential is a consequence of the
HFB approach, provided the usual nucleon--nucleon interaction
is enhanced on the nuclear surface where the nuclear density decreases.
The microscopic formalism to estimate the $\alpha$--particle formation probability
has been developed previously (see Refs. \cite{Del92,Del10}), 
but for the completeness of the overall presentation we will briefly describe those features 
which are of interest for the present work.

\section{Theoretical background} 
\label{sec:theor} 
\setcounter{equation}{0} 
\renewcommand{\theequation}{2.\arabic{equation}} 

\subsection{Surface Gaussian Interaction (SGI)}

The $\alpha$-decay process
\bea
P(\textrm{parent})\rightarrow D(\textrm{daughter})+\alpha
\eea
is allowed when the energy release (Q-value) is positive. This surplus is transformed into 
the relative kinetic energy of the $\alpha$--core system $Q=\mu_{\alpha}v^2/2$, where 
$\mu_{\alpha}$ is the reduced mass of the daughter--$\alpha$ system.
$\alpha$-decay between ground states (gs) takes place for select few
very light elements (for example, $^{5}$He, $^{5}$Li, $^8$Be) and becomes much more
prevalent in the region of the nuclear chart with $Z>50$.
The basic requirement to properly describe emission processes is that the basis 
wave functions follow a correct asymptotic behavior. It turns out that the asymptotic value
of sp orbitals provided by the standard Woods--Saxon potential is too
small to reproduce the experimental value of the $\alpha$-decay width.
A successful solution to this problem was proposed in Ref. \cite{Var92},
where the decaying state was described by a combination of a shell--model 
wave function  plus a cluster component $\Phi=\Phi_{SM}+\Phi_{\textrm{clus}}$.
The cluster component is expected to contain the high--lying shell 
model configurations, and the shell model component  $\Phi_{SM}$ is evaluated within a major shell only. 
The cluster component  $\Phi_{\textrm{clus}}$ is expanded in terms of shifted Gaussians and 
is used to diagonalize the residual two--body interaction.
A similar method was recently applied to describe anomalous large B(E1), B(E2) values and 
$\alpha$-decay half lives corresponding to transitions from states of $^{212}$Po \cite{Del12}. 

A different proposal was presented in Ref. \cite{Del13}, namely the use of a sum between a Woods--Saxon mean field 
and a Gaussian potential centered beyond the nuclear surface at
$R_{cl}=1.3~\left(A_D^{1/3}+4^{1/3}\right)$ with a length parameter $b_{cl}=1~{\rm fm}$.
This was used to generate sp orbitals able to properly describe the absolute value 
of $\alpha$-decay widths from even-even emitters.
A similar potential, but with a Woods--Saxon formfactor multiplied by a Gaussian clustering 
correction was used in Ref. \cite{Bai19} to describe $\alpha$-clustering in some emitters above doubly magic nuclei.

Various nuclear collective states are described within a microscopic formalism by a residual interaction
peaked on the nuclear surface. In particular, in this work we will describe two--particle 
($pp$, $nn$) collective states formed by a nucleon--nucleon residual interaction enhanced on the nuclear surface. 
In doing this, we generalize the well-known Surface Delta Interaction (SDI) in the form of the Surface Gaussian Interaction (SGI)
\bea
\label{SGI}
&&v_{SGI}({\bf r_{\tau},R_{\tau}})=v_{rel}(r_{\tau})v_{cm}(R_{\tau})
\nn&=&
-v_0\exp\left(-\frac{|{\bf r}_{\tau}|^2}{b^2_{rel}}\right)
\exp\left(-\frac{(|{\bf R}_{\tau}|-R_0)^2}{b^2_{cm}}\right)
\eea
given here in terms of the the relative and cm coordinates
${\bf r}={\bf r}_{1\tau}-{\bf r}_{2\tau}$, 
${\bf R}_{\tau}=({\bf r}_{1\tau}+{\bf r}_{2\tau})/2$.
We will add this component to the standard nucleon-nucleon interaction $v_{rel}(r_{\tau})$,
given by the usual Gaussian shape
\bea
\label{V}
v(r_{\tau},R_{\tau})&=&-v_0\exp\left(-\frac{r^2_{\tau}}{b^2_{rel}}\right)
\nn&\times&
\left[1+x_c\exp\left(-\frac{(R_{\tau}-R_0)^2}{b^2_{cm}}\right)\right],
\eea
where $r_{\tau}=|{\bf r}_{\tau}|$, $R_{\tau}=|{\bf R}_{\tau}|$ and
$x_c$ plays the role of the mixing residual strength, common for protons and neutrons.

\subsection{Cluster Hartree--Fock--Bogoliubov Approach}

The mean field can be generated by diagonalizing the HFB equations \cite{Rin80}
\bea
\label{HFB}
&&\left[-\frac{\hbar^2}{2\mu}\nabla^2+\Gamma^{(dir)}({\bf r})\right]\psi_{am}({\bf r})
\nn&+&
\int d{\bf r'}\Gamma^{(exc)}({\bf r,r'})\psi_{am}({\bf r'})
=\epsilon_a\psi_{am}({\bf r}),
\eea
depending upon direct and exchange potentials
\bea
\label{pot}
\Gamma^{(dir)}({\bf r}_{\tau})&=&\int d{\bf r}_{\tau}'v({\bf r}_{\tau},{\bf r}_{\tau}')\rho({\bf r}_{\tau}')
\nn
\Gamma^{(exc)}({\bf r}_{\tau},{\bf r}_{\tau}')&=&-v({\bf r}_{\tau},{\bf r}_{\tau}')\rho({\bf r}_{\tau},{\bf r}_{\tau}')
\nn
\tau&=&p,n
\eea
in terms of densities
\bea
\label{dens}
\rho({\bf r}_{\tau})&=&\sum_{a=1}^{n_{\tau}}V_{\tau a}^2\sum_{m=-j_a}^{j_a}\left|\psi_{am}({\bf r}_{\tau})\right|^2
\nn
\rho({\bf r}_{\tau}{\bf r}_{\tau}')&=&\sum_{a=1}^{n_{\tau}}V_{\tau a}^2
\sum_{m=-j_a}^{j_a}\psi^*_{am}({\bf r}_{\tau}')\psi_{am}({\bf r}_{\tau}).
\eea
We use the standard plus surface residual potential (\ref{V}) and we call this
procedure the Cluster HFB (CHFB) approach. This clustered mean field describes
the dynamics of proton and neutron quasiparticle pairs.
The amplitudes $U_{\tau a}$, $V_{\tau a}$ are given by the quasiparticle creation operator 
written in terms of the particle operators
\bea
\label{quas}
\alpha^{\dag}_{am_a}=U_{a}c^{\dag}_{am_a}+V_{a}c_{a-m_a}
\eea
where $a=(\tau_a\epsilon_a l_aj_a)$. They satisfy the standard system of gap equations
\bea
\label{gaps}
\Delta_a&=&\sum_{b=1}^{n_{\tau}}G_0(ab)\Omega_bU_bV_b=
v_0\sum_{b=1}^{n_{\tau}}G^{(0)}_0(ab)\frac{\Omega_b\Delta_b}{2E_b}
\nn
a&=&1,2,...,n_{\tau}
\eea
where $n_{\tau}$ is the number of considered sp levels and 
\bea
\Omega_b=\oh\what{j_b}^2=j_b+\oh.
\eea
The monopole pairing interaction is given by
\bea
G_0(ab)=-\frac{4}{\what{j_a}\what{j_b}}\la aa;0|v| bb;0\ra
=v_0G^{(0)}_0(ab).
\eea
In Appendix A we estimate the matrix elements of this interaction
for the wave functions provided by the diagonalization of the mean field.
The amplitudes
\bea
\left(\matrix{U_a\cr V_a}\right)=\frac{1}{\sqrt{2}}
\left(1\pm\frac{\epsilon_a-\lambda_{\tau}}{E_a}\right)^{\oh},~~~\tau=p,n
\eea
are defined in terms of the quasiparticle energy
\bea
E_{a}&=&\sqrt{(\epsilon_a-\lambda_{\tau})^2+\Delta_a^2},~~~\tau=p,n
\eea
where $\lambda_{\tau}$ are Lagrange multipliers accounting for the conservation of the number of particles.
We solve the system (\ref{gaps}) by looking for an effective strength of the pairing interaction $v_0$ 
required to obtain the experimental value of the gap parameter at the Fermi level.
It can be approximated by the well known ansatz
\bea
\Delta_{a_F}=\Delta_{exp}\sim\frac{12}{\sqrt{A}}~{\rm MeV}.
\eea
In Appendix B we show that the CHFB procedure predicts a mean field potential of the form
\bea
\label{vl}
V_{MF}(r_{\tau})&=&V_{0}(r_{\tau})+V_{cl}(r_{\tau}),~\tau=p,n.
\eea
$V_0$ describes the standard mean field close to the Woods--Saxon shape.
It has a somewhat involved expression following from computational details
that are not crucial for the physics of this discussion. These details are described
in Appendix B and the expression for the potential is given in Eq. ({\ref{vzvcl}})
in terms of other quantities defined and computed there. $V_{cl}$ is also
described in detail in the same appendix, but it can be written in Gaussian form 
\bea
\label{gauss}
V_{cl}(r_{\tau})&=&A_{cl} exp\left[-\left(\frac{r_{\tau}-R_{cl}}{b_{cl}}\right)^2\right].
\eea
The cluster parameters can be derived analitically for a step--function density in terms of original sp 
interaction parameters (\ref{V}), with a proof being outlined in Appendix B leading
to Eq. (\ref{rclsp}). $R_{0}$ is parametrized in Eq. (\ref{rtrz}). The length parameters $b_{rel}$, $b_{cm}$ and 
$b_{cl}$ characterize the corresponding Gaussians found in the structure of the potential
(\ref{vl}). Their values are once again discussed in Appendix B and shown to be those
in Eq. (\ref{rclsp}):
\bea
\label{cond}
 R_{cl}&=&R_0
 \nn
 b_{cl}&=&\sqrt{2}b_{cm}=b_{rel}/\sqrt{2}.
\eea
Our numerical analysis has shown that the realistic sp densities (\ref{dens}) provide results
that are very close to the above analytic approximations.
The inclusion of the SGI residual interaction in simultaneously solving the mean field (\ref{HFB}) and
pairing equations (\ref{gaps}) is a procedure going beyond the mean field approach \cite{Sch21}.
In our case it describes collective $pp$ and $nn$ pair states entering the structure of
the $\alpha$-particle. $pn$--pairing generally has a very small contribution 
to $\alpha$--decay from heavy nuclei \cite{Bar16} and is therefore neglected here.
Thus, we can justify on microscopic grounds the use of a similar potential in Ref. \cite{Del13}.

\subsection{Decay Width for Deformed Nuclei}

A very good approximation of the total decay width connecting the gs of deformed 
even-even nuclei is given by the following factorization \cite{Lan58,Del10,Ghi21}
\bea
\label{Gamma}
\Gamma=\Gamma_0D(\beta_2)
\eea
between the monopole decay width
\bea
\label{spher}
\Gamma_0=\hbar v\left[\frac{{R\cal F}_{0}(R)}{G_0(\chi,\rho)}\right]^2
\eea
where $R$ is the $\alpha$-core center of mass (cm) radius,
and the deformation factor 
\bea
\label{D}
D(\beta_2)&=&\sum_L\exp\left[-2\frac{L(L+1)}{\chi}\sqrt{\frac{\chi}{\rho}-1}\right]
{\cal K}^2_{L0}(\beta_2)
\nn
\eea
induced by the Coulomb field characterized by the quadrupole deformation $\beta_2$.
Here, $G_0(\chi,\rho)$ is the monopole irregular Coulomb function depending upon the Coulomb parameter
$\chi=4Z_De^2/(\hbar v)$ and reduced radius $\rho=\kappa R$, where $\hbar\kappa=\mu_{\alpha}v$ is the linear momentum and
\bea
{\cal K}_{LL'}(\beta_2)&=&\int_{-1}^1Y_{L0}(x)e^{\beta_2BP_2(x)}Y_{L'0}(x)dx
\nn
B&\equiv&\frac{2}{5}\chi\beta_2\left(2-\frac{\rho}{\chi}\right)\sqrt{\frac{5}{4\pi}\frac{\rho}{\chi}\left(1-\frac{\rho}{\chi}\right)}
\nn
\eea
defines the Fr\"oman propagator matrix \cite{Del10,Fro57}. Higher order 
multipoles of the nuclear shape are important in the description of
the $\alpha$-emission spectrum, particularly when transitions to excited
states are involved. However, one can still obtain good results when restricting
the analysis only to the quadrupole moment. For a more detailed discussion
and comparison of these methods, one can see Ref. \cite{Dum22} and references
indicated therein.

\subsection{Formation amplitude}

In the framework outlined above, the $\alpha$-particle formation amplitude can be calculated within a spherical approach.
It is given by the following overlap integral \cite{Del10}
\bea
\label{FR}
{\cal F}_0(R)=\la \Psi_P|\Psi_D\Psi_{\alpha}\ra
\eea
where $\Psi_P$, $\Psi_D$ and $\Psi_{\alpha}$ are the wave functions 
of the parent, daughter and $\alpha$--particle respectively. 
The above relation is a good approximation beyond the geometrical touching 
configuration, where the $\alpha$--core antisymmetrisation becomes less important.
It is convenient to write the formation amplitude  by using a harmonic oscillator 
(ho) representation since then all integrals can be performed analytically.
Thus, the wave function diagonalizing the mean field (MF) can be written
\bea
\psi_{\tau\epsilon ljm}(x)=\la x|\psi_{\tau\epsilon ljm}\ra=
{\cal R}_{\tau\epsilon lj}(r){\cal Y}^{(l{\oh})}_{jm}(\widehat{r},s)
\eea
where $x=({\bf r},s)$, in terms of the radial MF wave function and spin--orbit harmonics respectively
\bea
{\cal R}_{\tau\epsilon lj}(r)&=&\sum_nd_{\tau\epsilon lj}^{n}{\cal R}^{(\beta)}_{nl}(r)
\nn
{\cal Y}^{(l{\oh})}_{jm}(\widehat{r},s)&=&\left[i^lY_{l}(\widehat{r})\otimes\chi_{\oh}(s)\right].
\eea
Here ${\cal R}^{(\beta)}_{nl}(r)$ denotes the spherical ho wave function depending
upon the ho size parameter $\beta=M_N\omega/\hbar$. 
The formation amplitude becomes \cite{Del10}
\bea
\label{ampl}
{\cal F}_0(R)=\sum_{N_{\alpha}}{\cal W}_{N_{\alpha}}{\cal R}^{(4\beta)}_{N_{\alpha}0}(R)
\equiv \sum_{N_{\alpha}}{\cal F}_{N_{\alpha}0}(R)
\eea
where $N_{\alpha}$ is the ho radial quantum number corresponding to
the $\alpha$-particle motion with angular momentum $L_{\alpha}=0$.
The $\mathcal{W}$--coefficients are given by the following superposition
\bea
\label{WN}
&&{\cal W}_{N_{\alpha}}=8\sum_{n_{\alpha}N_pN_n}{\cal G}_{N_p}{\cal G}_{N_n}
\nn&\times&
\la n_{\alpha},0;N_{\alpha},0;0|N_p,0;N_n,0;0\ra
{\cal I}^{(\beta\beta_{\alpha})}_{n_{\alpha}0}
\eea
where the bra--ket product is the standard Talmi-Moshinky (TM) recoupling coefficient
connecting the $pp$ and $nn$ pairs to $\alpha$-particle coordinates.
Here, ${\cal I}$ is the overlap integral between the ho sp components ${\cal R}^{(\beta)}_{n_{\alpha}0}$
and the $\alpha$--particle wave function ${\cal R}^{(\beta_{\alpha})}_{00}$.
The quantity  ${\cal G}_{N_p}$ (${\cal G}_{N_n}$) contains only proton (neutron)
degrees of freedom
\bea
\label{GN}
{\cal G}_{N_{\tau}}&=&\sum_{n_1n_2lj}{\cal B}_{\tau}(n_1ljn_2lj;0)
\nn&\times&
\la\left(ll\right)0\left(\oh\oh\right)0;0|\left(l\oh\right)j\left(l\oh\right)j;0\ra
\nn&\times&
\sum_{n_{\tau}}\la n_{\tau}0N_{\tau}0;0n_1ln_2l;0\ra{\cal I}^{(\beta\beta_{\alpha})}_{n_{\tau}0},
\eea
where the bra--ket in the second line denotes the jj--LS recoupling coefficient and the $\mathcal{B}$--coefficient contains 
the nuclear structure information
\bea
\label{Btau}
{\cal B}_{\tau}(n_1ljn_2lj;0)=\frac{\what{j}}{\sqrt{2}}U_{\tau\epsilon lj}V_{\tau\epsilon lj}
d^{n_1}_{\tau\epsilon lj}d^{n_2}_{\tau\epsilon lj}.
\eea
Eq. (\ref{WN}) contains products of quantities which depend only on proton or
neutron degrees of freedom. 

\section{Numerical application} 
\label{sec:numer} 
\setcounter{equation}{0} 
\renewcommand{\theequation}{3.\arabic{equation}} 

The formation of an $\alpha$-cluster is a collective process, less sensitive to specific details
connected to the sp level structure. It turns out that the essential 
part of the sp mean field for decay processes is given by distances beyond the geometrical touching 
radius 
\bea\label{geotr}
R_c=1.2\left(A_D^{1/3}+A_{\alpha}^{1/3}\right).
\eea

\subsection{Mean field shape}

\begin{figure}
\begin{center} 
\includegraphics[width=9cm]{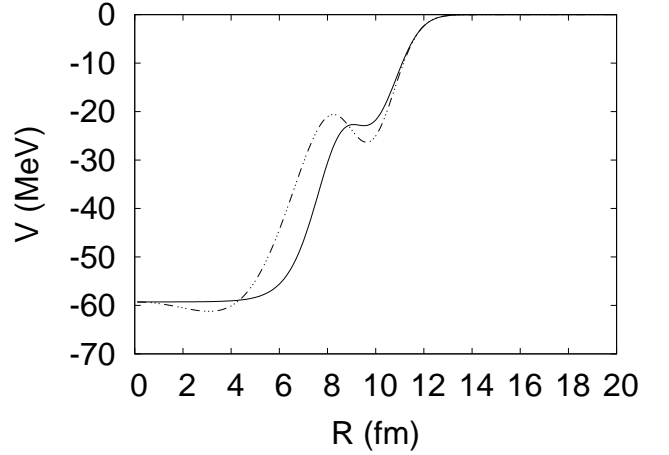} 
\caption{
Proton HFB mean field plus SGI interaction (dashed line) and
WS plus SGI potential (solid line) in the case of $^{242}\textrm{Pu}$. 
}
\label{fig1}
\end{center} 
\end{figure}

In Fig. \ref{fig1} we plotted the proton CHFB potential of Eq. (\ref{vl}) calculated for $^{242}\textrm{Pu}$ 
(dashed line) and Woods--Saxon potential with universal parameterisation \cite{Cwi78,Dud81,Dud82} 
plus SGI residual interaction (solid line), satisfying the conditions (\ref{cond}).
The residual strength $x_{c}$ has the value of $19~\textrm{MeV}$ which reproduces the observed
$\alpha$--decay width. The overall effect obtained is the formation of pocket--like
potential structures centered on the nuclear surface which favor nucleon clustering.
One notices that both versions give practically the same results concerning the estimate of the 
decay width beyond the geometrical touching radius $R_c$=9.38 fm.

\begin{figure}[h]
\begin{center} 
\includegraphics[width=9cm]{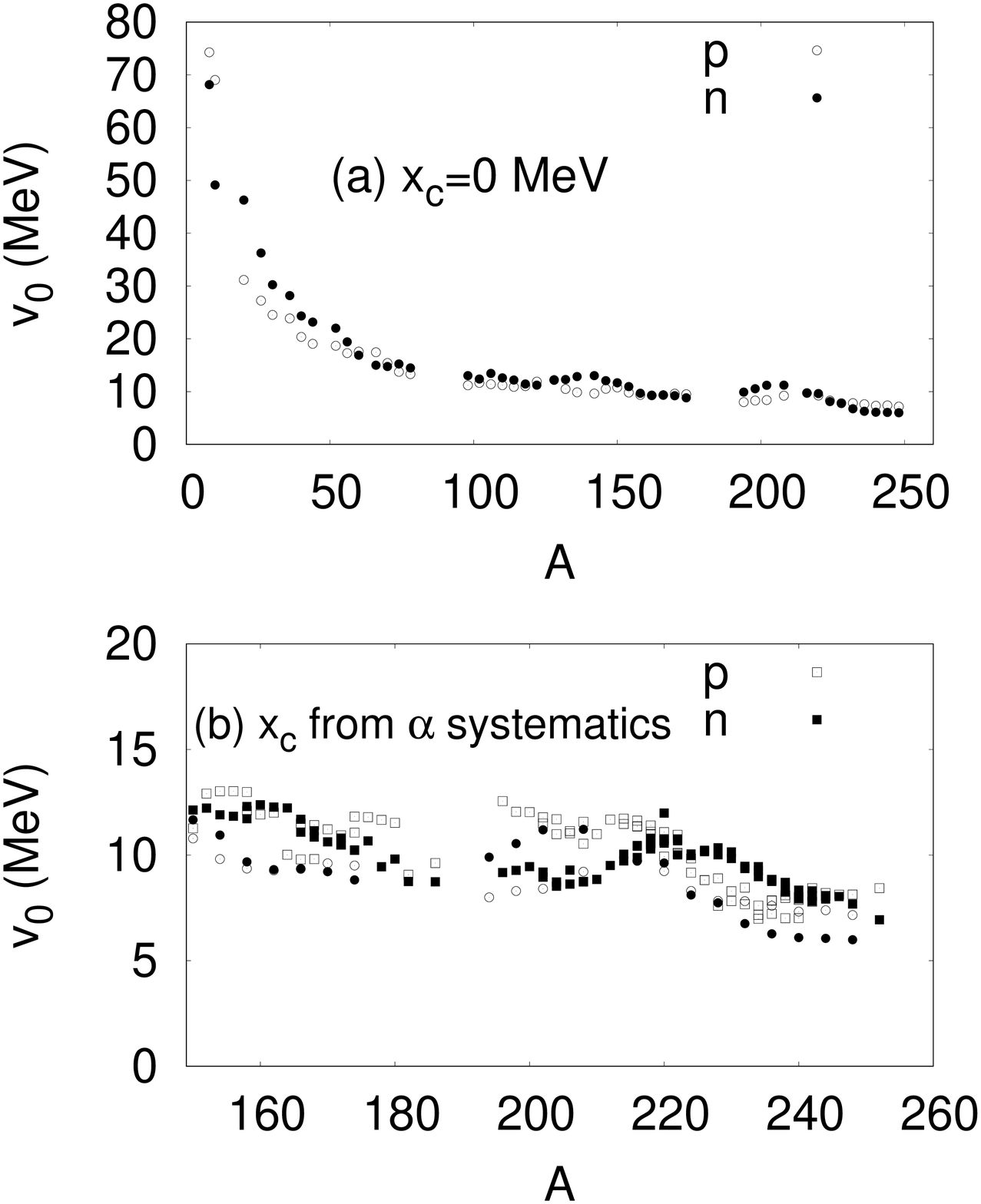} 
\caption{
Panel (a) shows the pairing interaction strength versus mass number across the nuclear chart
for $x_{c}=0$ (panel a). Panel (b) compares the case above (circles) with that of $\alpha$--emitters (squares) having $x_c$
taken from the decay systematics.
}
\label{fig2}
\end{center} 
\end{figure}

For this reason we performed our analysis by using a Woods--Saxon sp potential
with universal parameterisation plus a residual SGI, satisfying the conditions
(\ref{cond}) predicted by the CHFB formalism.
We considered the standard value of the nucleon-nucleon radius $b_{rel}$=2 fm
and a slighly larger radius than the touching radius $R_0=1.275(A_D^{1/3}+A_{\alpha}^{1/3})$,
corresponding to a small percent of the equilibrium nuclear density, as predicted by the
nuclear matter calculations of the $\alpha$-clustering transition. This value is known as the Mott density 
for the $\alpha$--formation \cite{Rop98,Toh17}.
Thus, the only free parameter of the model is the strength $x_c$ of the SGI and it was adjusted 
to reproduce experimental decay widths.

\subsection{Pairing strength systematics}

We analyzed superfluid even--even $\alpha$-emitters ranging from rare earths to 
actinides and superheavy nuclei with experimental data available at the ENSDF \cite{ENSDF}.

The main nuclear structure ingredients enter the $\mathcal{B}$-coefficients (\ref{Btau}). They
are given by the expansion coefficients of the sp orbitals in terms of ho components and
BCS amplitudes depending upon the strength of the pairing interaction.
Therefore, we began with the analysis this strength $v_0$ by using the systematics of the pairing gap.
Panel (a) of Fig. \ref{fig2} shows the pairing interaction strength versus the mass number across
the nuclear chart for the case of no residual interaction ($x_{c}=0$). 
Similarly, panel (b) shows the same plot compared with the case of $\alpha$--emitters 
having their values of $x_{c}$ taken from the decay systematics. What is observed
in the first case is a significant increase of the pairing strength for small mass numbers.
This behavior is consistent with a recent microscopic description of two-proton emitters \cite{Ghi22},
where a value $v_0\sim$ 45 MeV was obtained in free space in order to reproduce 
the experimental value of a simultaneous two--proton decay width.
Notice that the mean value for $\alpha$--emitters with A$>$150 is of $\approx 9~\textrm{MeV}$ 
for the $pp$ and $nn$ pairing strengths respectively.
Turning on the residual interaction, these values go to roughly $~\approx 10~\textrm{MeV}$,
so they do not change significantly. 

\subsection{Analysis of the plateau condition}

\begin{figure}[h]
\begin{center} 
\includegraphics[width=9cm]{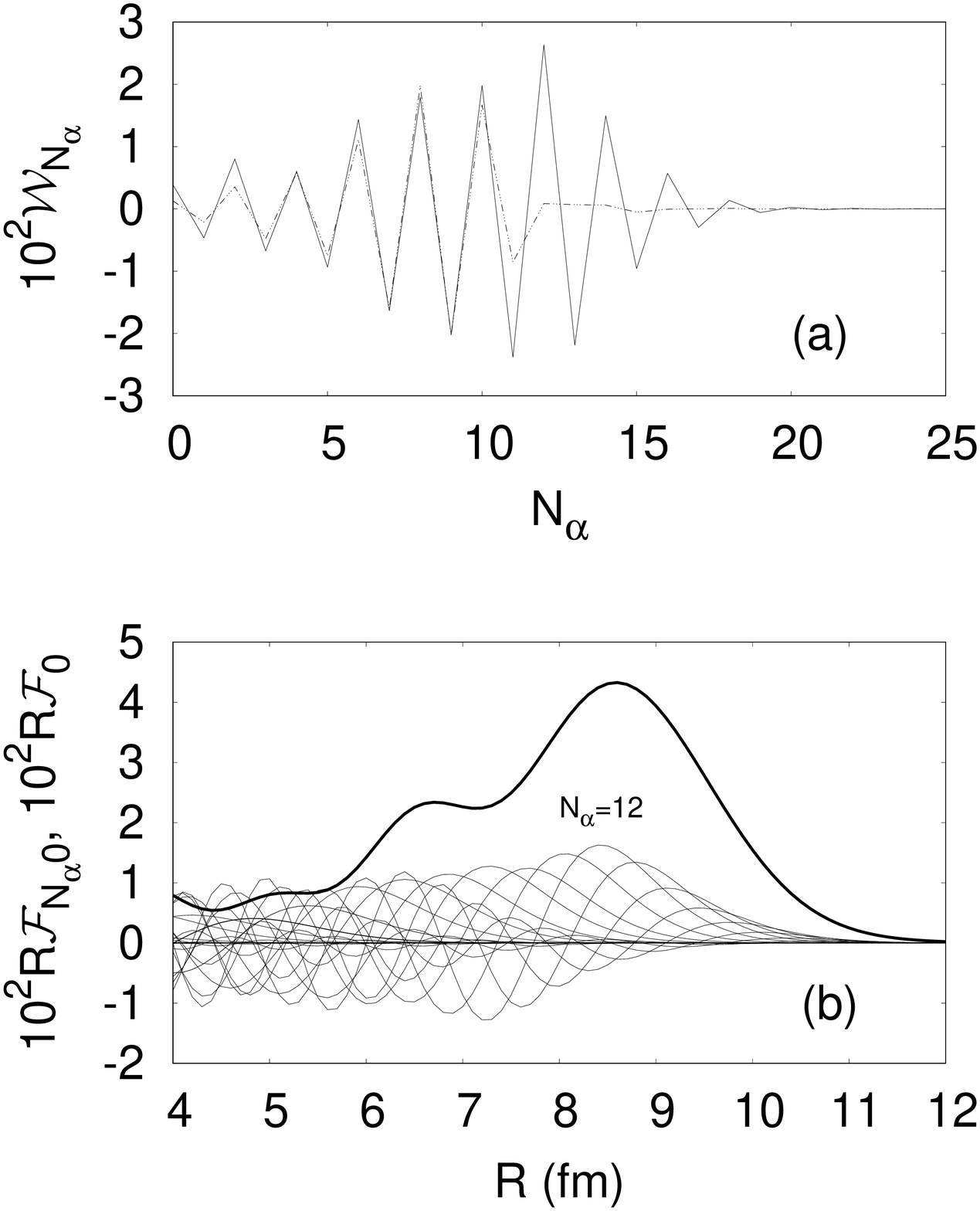} 
\caption{
(a) $\mathcal{W}$-coefficients (\ref{WN}) versus the quartet radial quantum number $N_{\alpha}$
in the absence of SGI interaction (dot--dashed line) and for $x_c$=19 MeV (solid line)
corresponding to the decay of $^{242}\textrm{Pu}$.
(b) The radial components of the $\alpha$--formation amplitude (\ref{ampl})
(thin solid lines) and the total value (thick solid line).
}
\label{fig3}
\end{center} 
\end{figure}

As we already mentioned, the clustering process takes place on the nuclear surface, where the low density
favores the formation of $\alpha$-particles. According to Eq. (\ref{ampl})
the formation amplitude is a coherent superposition of four-body radial ho functions
multiplied by $\mathcal{W}$-coefficients, plotted in the panel (a) of Fig. \ref{fig3} for the decay
of $^{242}\textrm{Pu}$. By a dot--dashed line are given the $\mathcal{W}$--coefficients corresponding
to the absence of the residual SGI interaction ($x_c=0$), while the solid line denotes
the case reproducing the experimental decay width, namely $x_c$=19 MeV. One notices the
occurence of large components with $N_{\alpha}>$ 10 in the latter case. 
In spite of the staggered character of these coefficients, the products with ho functions
${\cal F}_{N_{\alpha}0}(R)$ plotted in panel (b) have a coherent behavior.
They give the maximum of the summed formation amplitude ${\cal F}_0(R)$, plotted in the
same panel by a thicker line. Its maximal value corresponds to the larger
component with the cm radial quantum number $N_{\alpha}$=12.

\begin{figure}[h]
\begin{center} 
\includegraphics[width=9cm]{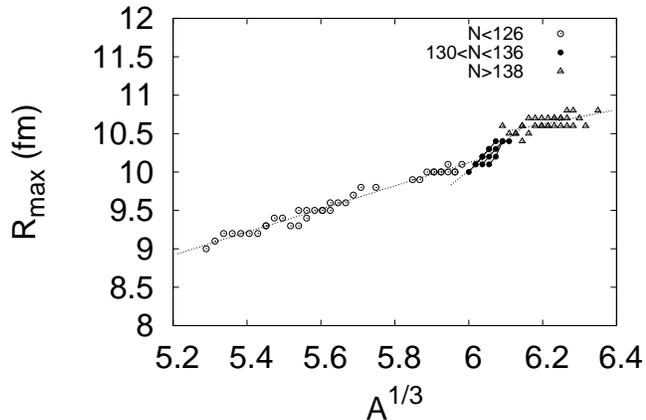} 
\caption{
Radius corresponding to the peak of the $\alpha$--particle wavefunction versus
parent mass number to the power $\frac{1}{3}$.
}
\label{fig4}
\end{center} 
\end{figure}

Fig. \ref{fig4} shows the systematics for the radius corresponding to the 
maximal value of the $\alpha$--particle formation amplitude versus the parent mass number
to the power $\frac{1}{3}$. One observes three regions of linear correlations,
corresponding to the neutron numbers $N<126$ (empty circles), $130\le N\le 136$ (filled circles) 
and $N\ge 138$ (empty triangles).

\begin{figure}[h]
\begin{center} 
\includegraphics[width=9cm]{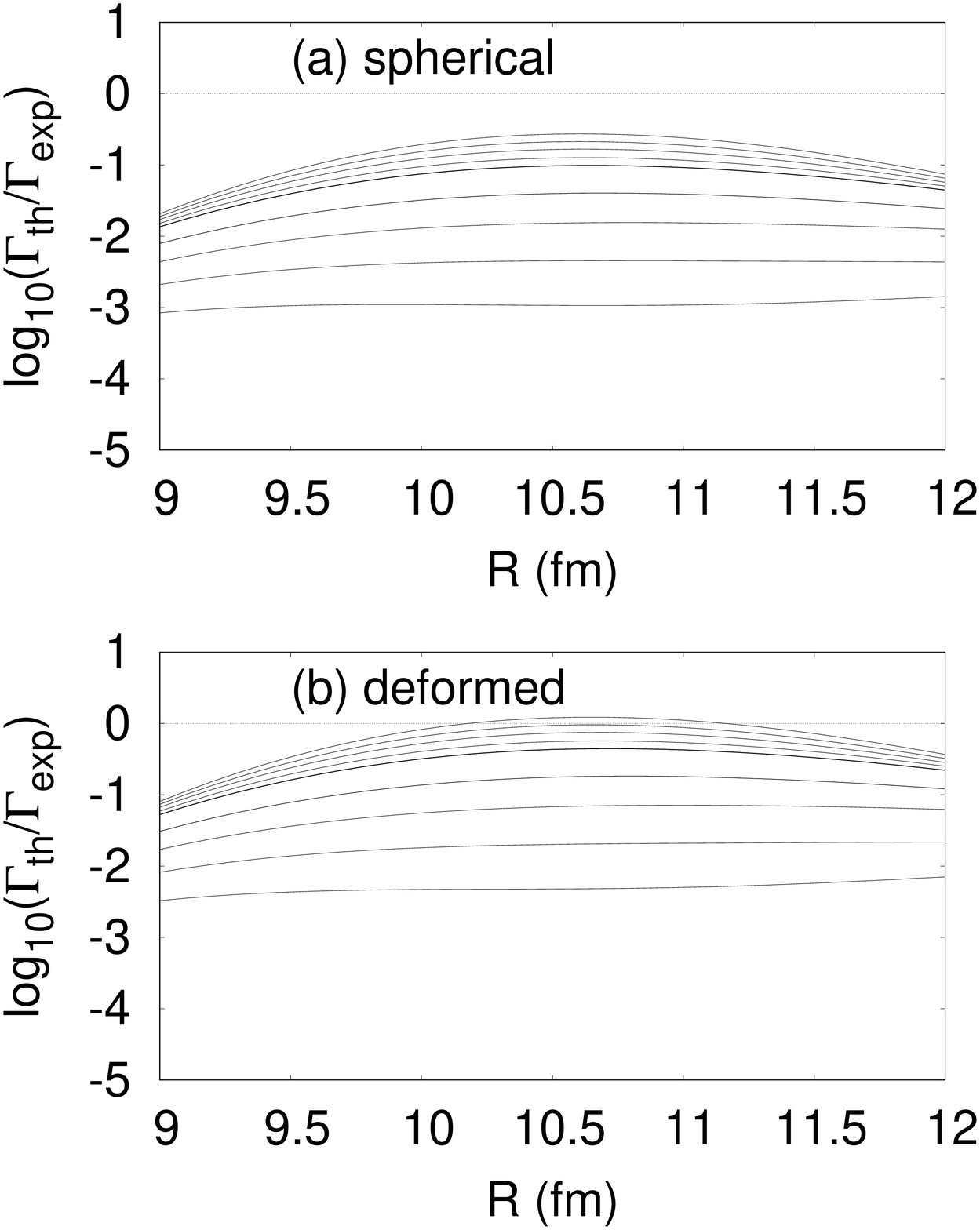} 
\caption{
Logarithm of the ratio between the theoretical and experimental decay width
versus radius in the case of spherical (panel a) and deformed (banel b)
calculations. The nucleus is $^{242}\textrm{Pu}$ and $x_{c}$ ranges between
$5-19~\textrm{MeV}$, with the smaller values corresponding to a wider plateau.
}
\label{fig5}
\end{center} 
\end{figure}

\begin{center}
\begin{table}[h]
{\bf Table 1}
{\it Systematics of peak radius versus mass number to the power $\frac{1}{3}$}
\vskip5mm
\begin{tabular}{|c|c|c|c|}
\hline
region & $a$ & $b$ & $\sigma$ \cr
\hline
$N<126$            & 1.503 &   1.102 & 0.052 \cr 
$130\le N\le 136$  & 3.850 & -13.082 & 0.060 \cr 
$N\ge 138$         & 0.932 &   4.845 & 0.072 \cr 
\hline
\end{tabular}
\end{table}
\end{center}

The first and third regions are in fact quite similar in behavior, with the second
region bridging them. The separation between the second and third regions becomes
unambigous if one looks at Fig. \ref{fig6} to be discussed in section III.4.  
It is interesting to observe that the second region is comprised of
Rn, Ra, Th and U isotopes, the lightest one being $^{216}\textrm{Rn}$ while the
heaviest nucleus is $^{228}\textrm{U}$. These two configurations of nucleons can
be imagined as a $^{208}\textrm{Pb}$ core coupled to a number of $\alpha$--particles
of 2 and 5 respectively, with all other nuclei in between having a number of nucleons
compatible with arrangements consisting of a $^{208}\textrm{Pb}$ core, a number of 2--4 $\alpha$--particles
and an additional number of 1--3 $pp$ or $nn$ pairs. We are not stating that
this is indeed an accurate physical picture, but it does tie further into
the discussion of section III.4 and Fig. \ref{fig7}, where the data pertaining to this region
suggests enhanced clustering features due to the small number of nucleons found
above the closed shells of $^{208}$Pb. In any case, the slope, intercept and
standard deviation following for a basic linear fit of the data for each region are given in Table 1.

The calculated decay width (\ref{Gamma}) should not depend upon the radius beyond the
nuclear surface, thus satisfying the so-called plateau condition, due to the fact that 
in a phenomenological appropach both internal $R{\cal F}_0$ and external $G_0(R)$ functions 
satisfy the same Schr\"odinger equation. 
Our case is that of a semi--microscopic approach. The internal formation amplitude in (\ref{spher}) is provided 
by a microscopic method, while the external wave function satisfies the Coulomb equation and 
therefore the plateau condition is not automatically satisfied.

In order to check to what extent the plateau condition is satisfied we analyzed
the behavior of the calculated decay width for different values of the residual strength.
The result is shown in Fig. \ref{fig5} as a function of radius in the case of the parent
nucleus $^{242}\textrm{Pu}$. 
The results are shown for two different types of calculations. Panel (a) 
is for the computation without the Fr\"{o}man correction, while panel (b) shows the results corrected
for the nuclear deformation within the Fr\"{o}man approximation. 
$x_{c}$ ranges between $5-19~\textrm{MeV}$
with smaller values corresponding to broader plateaus in the logarithm of the 
decay widths ratio. One observes that the theoretical calculations converge to the observed value
with increasing values of $x_{c}$. Furthermore, the calculations corrected for nuclear deformation make a better estimate
of the decay width by a factor of roughly $5$ over the spherical calculation
for a given value of the residual strength. This underlines once again the
importance of nuclear deformation in the barrier penetration process. The approximate plateau 
condition is established at a little over $10~\textrm{fm}$, that is about
$1~\textrm{fm}$ beyond the geometrical contact radius. It is important to stress that we determined the strength $x_c$ 
reproducing the experimental decay width by using the following condition
\bea
\Big\la \log_{10}\frac{\Gamma^{def}_{th}(R)}{\Gamma_{exp}} \Big\ra=0
\eea 
where the mean value is considered in the interval of $\pm$1 fm around the radius $R_{max}$
where the maximal value is reached.

\subsection{Decay width systematics}

\begin{figure}[h]
\begin{center} 
\includegraphics[width=9cm]{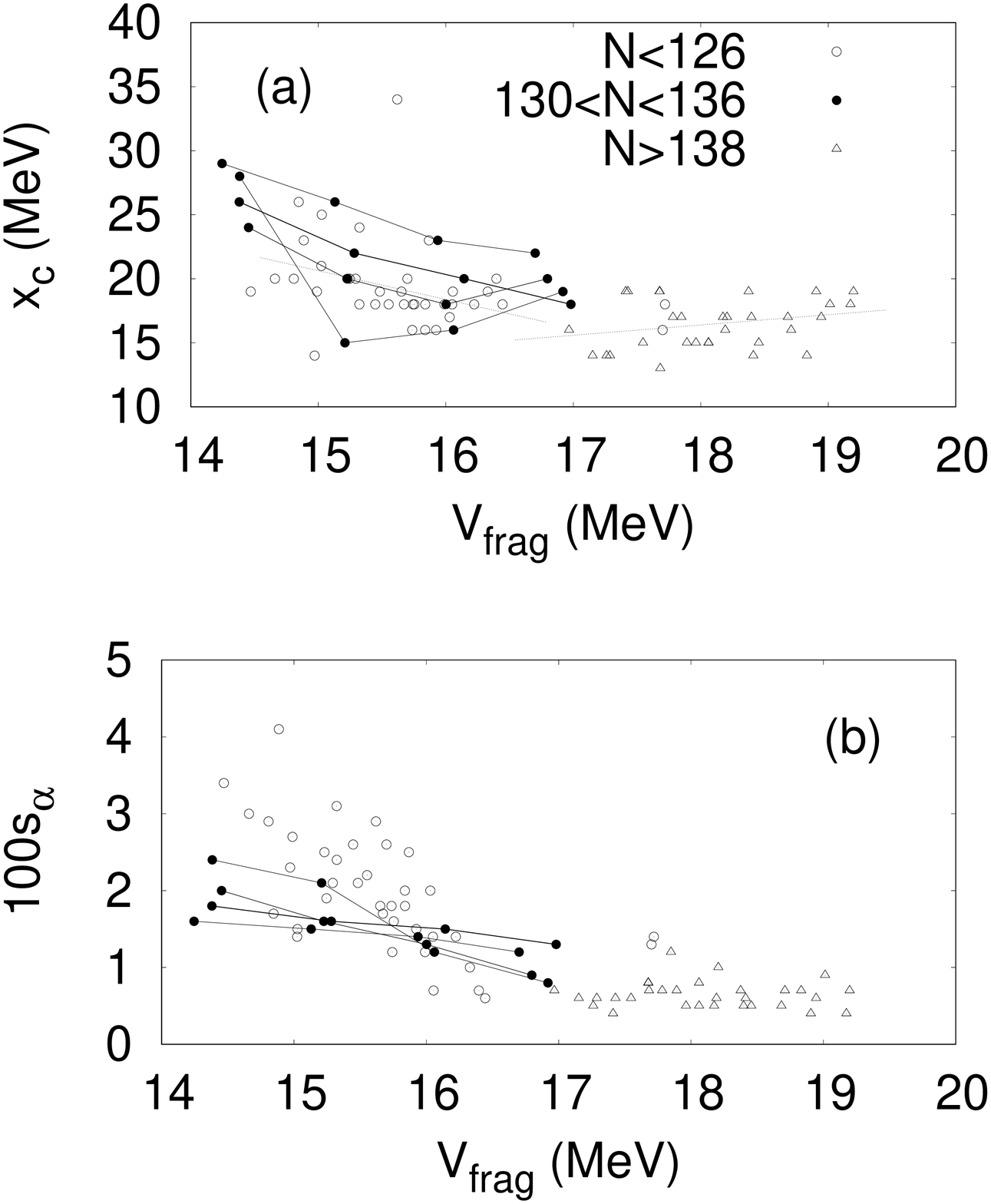} 
\caption{
Residual interaction strength (panel a) and $\alpha$--decay spectroscopic
factor (panel b) versus the fragmentation potential. 
}
\label{fig6}
\end{center} 
\end{figure}

Systematic calculations of $\alpha$--decay widths are presented in Fig. \ref{fig6},
namely the dependence of the residual interaction strength (panel a) and
spectroscopic factor (panel b) 
\bea
s_{\alpha}=\int_0^{\infty}|R{\cal F}_0(R)|^2dR,
\eea
on the fragmentation potential as suggested by the phenomenological systematics of Ref. \cite{Dum22}.
One observes once again two major trends with a transition region in the same
neutron ranges as found in the systematics of Fig. \ref{fig4}. For the first
and third regions, the slope, intercept and standard deviation are given in Table 2. 

\begin{center}
\begin{table}[h]
{\bf Table 2}
{\it Systematics of the residual interaction strength versus fragmentation potential}
\vskip5mm
\begin{tabular}{|c|c|c|c|}
\hline
region & $a$ & $b$ & $\sigma$ \cr
\hline
$N<126$            & -2.248 &  54.351 & 2.070 \cr 
$N\ge 138$         &  0.804 &   1.919 & 1.762 \cr 
\hline
\end{tabular}
\end{table}
\end{center}

\begin{figure}[h]
\begin{center} 
\includegraphics[width=9cm]{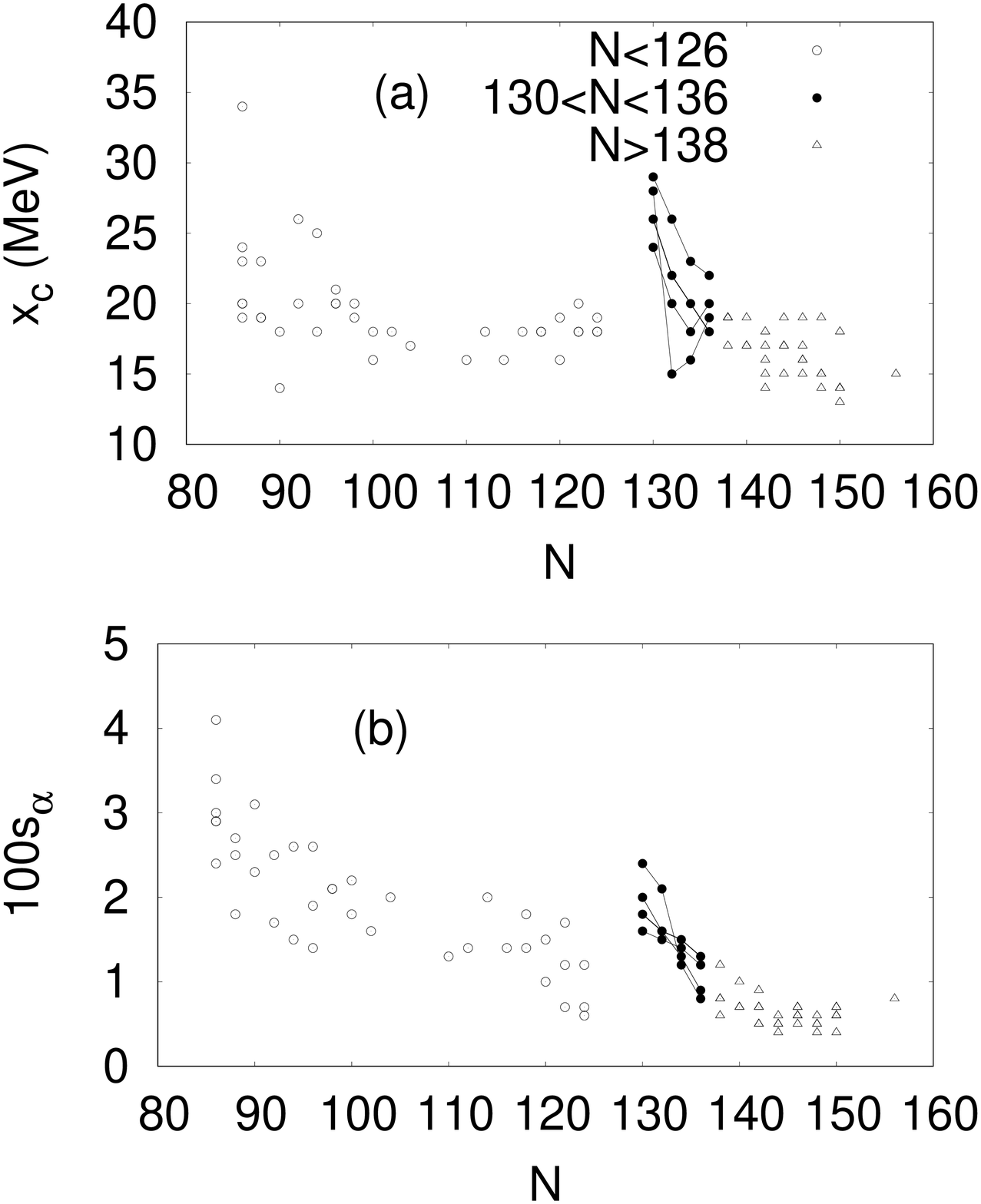} 
\caption{
Residual interaction strength (panel a) and $\alpha$--decay spectroscopic
factor (panel b) versus the neutron number. 
}
\label{fig7}
\end{center} 
\end{figure}

It is interesting to note that this phenomenon is reminiscent of a very similar feature found in 
proton--emission. There, the proton--decay spectroscopic factor exhibits
two trends around the charge number $Z=68$ where both
shape--coexistence phenomena and an abrupt change from oblate to prolate
deformations are observed \cite{Del21,Dum22}. 
However, as noted previously, in the case of $\alpha$--decay clustering phenomena play
a very important role in the dynamics of this particular transition. This is seen
in Fig. \ref{fig7} where the residual interaction strength (panel a) and spectroscopic
factor (panel b) are plotted versus the neutron number. One observes the typical
behavior of large clustering near closed shells followed by a decreasing trend. 

\begin{center}
\begin{table}[h]
{\bf Table 3}
{\it Predictions for superheavy even--even $\alpha$-emitters. Deformation
parameters are taken from \cite{Mol95}. Uncertainties relative to the recommended
value of the total half-life are taken from the maximal values tabulated in Ref. \cite{ENSDF}
at the time of this writing.}
\vskip5mm
\begin{tabular}{c c c c c c c c}
\hline
n & Nucleus &  $\beta_{2}$ & $ Q $       & $V_{\textrm{frag}}$       & $\log_{10}\Gamma_{\textrm{exp}}$   & $\log_{10}\frac{\Gamma_{\textrm{th}}}{\Gamma_{\textrm{exp}}}$ & $\epsilon$ \cr
  &         &              & MeV         & MeV                       & MeV                           & & \% \cr
\hline
 1 & $^{266}_{106}$Sg & 0.230 & 8.762 {\it 51} & 17.603 &-23.420 & 1.024 & 95 \cr
 2 & $^{264}_{108}$Hs & 0.229 &10.591 {\it 20} & 16.332 &-18.545 & 0.242 & - \cr
 3 & $^{266}_{108}$Hs & 0.230 &10.335 {\it 20} & 16.537 &-18.703 &-0.168 & 5 \cr
 4 & $^{270}_{108}$Hs & 0.231 & 9.300 {\it  7} & 17.470 &-21.896 & 0.435 & 6 \cr
 5 & $^{270}_{110}$Ds & 0.221 &11.200 {\it 50} & 16.075 &-17.341 &-0.232 & 35 \cr
 6 & $^{286}_{114}$Fl &-0.096 &10.345 {\it 60} & 17.528 &-20.943 &-0.003 & 24 \cr
 7 & $^{288}_{114}$Fl & 0.053 &10.090 {\it 70} & 17.733 &-21.244 &-0.420 & 22 \cr
 8 & $^{290}_{116}$Lv & 0.072 &11.000 {\it 80} & 17.270 &-19.517 &-0.234 & 4.2 \cr
 9 & $^{292}_{116}$Lv &-0.070 &10.800 {\it 70} & 17.420 &-19.597 &-0.658 & 5 \cr
10 & $^{294}_{118}$Og &-0.087 &11.810 {\it 60} & 16.855 &-18.596 & 0.172 & 76 \cr
\hline
\end{tabular}
\end{table}
\end{center}

In phenomenological studies of the $\alpha$--spectrum fine structure using a 
monopole plus quadrupole--quadrupole (QQ) interaction, the coupling strength of the QQ component
behaves in an analogous manner and is proportional to the reduced width, thereby acting 
as a measure of clustering on the nuclear surface \cite{Del13a}.

\begin{figure}[h]
\begin{center} 
\includegraphics[width=9cm]{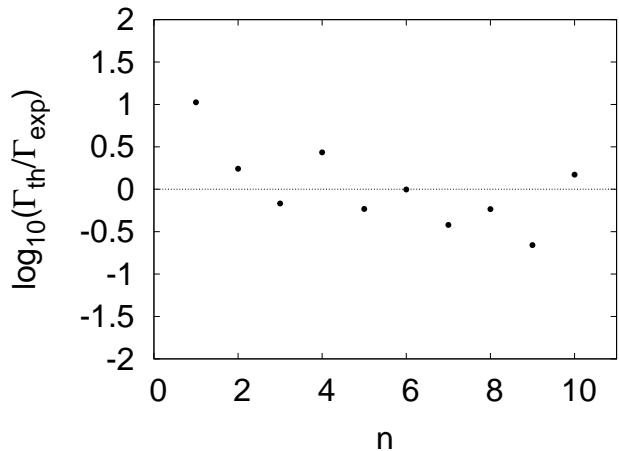} 
\caption{
Logarithm of the ratio for the predicted to experimental decay width
versus index number for even--even superheavy $\alpha$--emitters. 
}
\label{fig8}
\end{center} 
\end{figure}

\subsection{Predictions for superheavy emitters}

In order to test the predictive power of the model, we have used the systematics
of Table 2 to calculate the decay widths of known even--even superheavy emitters. 
The results are shown in Fig. \ref{fig8}, namely the logarithm of the ratio between
the calculated and experimental widths function of the index number of Table 3.
In spite of the somewhat large scattering of data for actinides in the range $N\ge 138$,
one observes an overall good agreement between the calculated and experimental values
for the decay widths of superheavy emitters, usually within a factor of 3. This is
quite reasonable in the context of the experimental uncertantities involved in these
measurements. The last column of Table 3 contains the quantity $\epsilon$, namely
the ratio of the largest recorded uncertainty in the total measured half-life relative
to the recommended value tabulated in Ref. \cite{ENSDF} at the time of this work. 
In contrast, similar experimental uncertainties in the region of the actinides 
where the relevant data are fitted tend to be smaller, of the order of $\approx 1\%$ or less.

Of particular interest is the case of the parent nucleus $^{266}_{106}$Sg. Not only
are the reported uncertainties in the total half-life quite large, but the
$\alpha$-decay branching ratio itself is currently recommended only as an
estimated lower bound of $\%\alpha\ge 18.0$. Perhaps the order of magnitude
discrepancy between the $\alpha$-decay width following from these reported values 
and our calculation is indicative of a measurement that can be improved. 

\section{Conclusions} 
\label{sec:concl} 
\setcounter{equation}{0} 
\renewcommand{\theequation}{4.\arabic{equation}} 

We have used the HFB mean field plus a residual nucleon--nucleon SGI in order to describe
$\alpha$-clustering in even-even nuclei.
We call this method the Cluster HFB (CHFB) approach. 

We have shown that the shape of the resulting mean field is close to the Woods--Saxon
potential with universal parameterisation plus a Gaussian clustering correction
with parameters determined by the residual nucleon-nucleon SGI. The strength of the residual interaction
was chosen to reproduce experimentally observed decay widths. We have shown
that the residual strength evaluated in this way is linearly correlated with
the fragmentation potential which is in agreement with the behavior of the
$\alpha$--particle preformation inferred from phenomenological theories. 
We have evidenced two such regions of linear correlation for emitters in the
range between rare earths and actinides, the transition between the two regimes
corresponding to the well-known high clustering found in the region above $^{208}$Pb.
The predictive power of the model was tested by estimating the half-lives of
superheavy $\alpha$--emitters, with good agreement being found with the experimental widths.

\section*{Acknowlegments}

This work was supported by the grant of the Romanian Ministry of Education and Research No. PN-19060101/2019-2022
and by the grant of the Institute of Atomic Physics
from the National Research -- Development and Innovation Plan III for 2015 - 2020/Programme 5/Subprogramme
5.1 ELI-RO, project ELI-RO No 12/2020.

\newpage
\section*{A. Matrix elements of the nucleon-nucleon interaction} 
\label{sec:appendA} 
\setcounter{equation}{0} 
\renewcommand{\theequation}{A.\arabic{equation}} 

The pairing function is given by
\bea
\Psi_{abJM}(x_1,x_2)&=&\delta_{ab}\delta_{J0}\delta_{M0}\left[\psi_a(x_1)\otimes\psi_a(x_2)\right]_{00}.
\nn
\eea
We first expand each sp wave function in terms of ho components
\bea
\Psi_{aa00}(x_1,x_2)&=&
\sum_{n_an'_a}d_{a}^{n_a}d_{a}^{n'_a}\Phi^{n_an'_a}_{aa00}(x_1,x_2)
\nn
\Phi^{n_an'_a}_{aa00}(x_1,x_2)&\equiv&\Big[
\left(\phi^{(\beta)}_{n_al_a}({\bf r}_1)\otimes\chi_{\oh}(s_1)\right)_{j_a}
\nn&\otimes&
\left(\phi^{(\beta)}_{n'_al_a}({\bf r}_2)\otimes\chi_{\oh}(s_2)\right)_{j_a}\Big]_{00}
\eea
and then we change from the $jj$ to the $LS$ coupling schere where one considers the spin singlet component.
Finally we change the radial part by using the Talmi--Moshinsky transformation
from absolute to relative and cm coordinates through the notation
$|\Phi^{n_an'_a}_{aa00}\ra\equiv|n_al_aj_an'_al_aj_a\ra$
\begin{widetext}
\bea
&&\la\Psi_{aa;0}(x_1,x_2)|v({\bf r,R})|\Psi_{bb;0}(x_1,x_2)\ra\equiv\la aa;0|V| bb;0\ra=
\sum_{n_an'_an_bn'_b}d_a^{n_a}d_a^{n'_a}d_b^{n_b}d_b^{n'_b}
\la n_al_aj_an'_al_aj_a|v({\bf r,R})|n_bl_bj_bn'_bl_bj_b\ra,
\nn\nn
&&\la n_al_aj_an'_al_aj_a|v({\bf r,R})|n_bl_bj_bn'_bl_bj_b\ra
\equiv
\la (l_al_a)0(\oh\oh)0;0 |(l_a\oh)j_a(l_a\oh)j_a;0\ra
\la (l_bl_b)0(\oh\oh)0;0 |(l_b\oh)j_b(l_b\oh)j_b;0\ra
\nn
&\times&
\sum_{lL}
\sum_{N}\la nlNL;0|n_al_an'_al_a;0\ra
\sum_{N'}\la n'lN'L;0|n_bl_bn'_bl_b;0\ra
\la{\cal R}^{(\beta/2)}_{nl}(r)|v_{rel}(r)|{\cal R}^{(\beta/2)}_{n'l}(r)\ra
\la{\cal R}^{(2\beta)}_{NL}(R)|v_{cm}(R)|{\cal R}^{(2\beta)}_{N'L}(R)\ra
\nn
\eea
\end{widetext}
where
\bea
2(n_a+n'_a+l_a)&=&2(n+N)+l+L
\nn
2(n_b+n'_b+l_b)&=&2(n'+N')+l+L.
\eea
For a potential depending only on the relative coordinate like the spin singlet gaussian interaction
\bea
v_{rel}(r)&=&-v_0\exp\left(-\frac{r^2}{b_{rel}^2}\right)~
\eea
the main building block becomes diagonal in $N$.

\section*{B. Mean field potential}
\label{sec:appendB} 
\setcounter{equation}{0} 
\renewcommand{\theequation}{B.\arabic{equation}} 

We calculate the direct and exchange potentials (\ref{pot}) depending on the densities (\ref{dens}).
As we have already shown, the spherical approach is accurate enough for the evaluation of the
$\alpha$--particle formation amplitude. Therefore the first density in (\ref{dens}) 
can be estimated in terms of the spherical sp wave functions summed on spin projections 
\begin{widetext}
\bea
\left|\psi_{a}({\bf r})\right|^2&=&\sum_{m=-1}^{j}\left|\psi_{am}({\bf r})\right|^2
={\cal R}^2_{\tau\epsilon lj}(r)
\sum_{m=-j}^{j}\left[{\cal Y}^{(l\oh)}_{jm}(\what{r},s)\right]^{\dag}{\cal Y}^{(l\oh)}_{jm}(\what{r},s)
\nn&=&
\frac{1}{4\pi}{\cal R}^2_{\tau\epsilon lj}(r)\left[(2j+1)+\sum_{L>0}^{2j}(2L+1)
\sum_{m=-j}^{j}C^{jLj}_{m0m}C^{jLj}_{\oh 0\oh}P_L(\what{r})\right]
\eea
\end{widetext}
and satisfying the normalisation rule
\bea
\int \left|\psi_{a}({\bf r})\right|^2d{\bf r}=2j+1.
\eea
As such, the density can be expanded as follows
\bea
\label{rho}
\rho(r,\cos\theta)&=&\rho_{0}(r)+\sum_{L>0}^{2j}\rho_{L}(r)P_L(\cos\theta)
\nn
\rho_{0}(r)&\equiv&\frac{1}{4\pi}\sum_a(2j_a+1)V^2_a{\cal R}^2_a(r)
\nn
\rho_{L}(r)&\equiv&\frac{1}{4\pi}\sum_aV^2_a{\cal R}^2_a(r)
\sum_{L>0}^{2j_a}(2L+1)C^{j_aLj_a}_{\oh 0\oh}
\nn&\times&
\sum_{m=-j_a}^{j_a}C^{j_aLj_a}_{m0m}.
\eea
Notice that the direct part of the potential with ${\bf r}\equiv{\bf r}_{\tau}$ is evaluated
\bea
\label{dir}
&&\Gamma^{(dir)}({\bf r})=\int d{\bf r}'v({\bf r},{\bf r}')\rho({\bf r}')
\nn&=&
V_{MF}(r,b_{rel},\infty,0)+x_cV_{MF}(r,b_{rel},b_{cm},R_0)
\nn
\eea
as a sum of two terms, namely a standard mean field potential
given by the relative inter-nucleon interaction
and a term given by the SGI inter--nucleon interaction (\ref{SGI}).
The general expresion of the mean field is obtained through the following integral, where
the major contribution is due to the monopole density term
\begin{widetext}
\bea
\label{VMF}
V_{MF}(r,b_{rel},b_{cm},R_0)&=&
-v_0\exp\left[-\left(\frac{r}{b_{rel}}\right)^2-\left(\frac{r-2R_0}{2b_{cm}}\right)^2\right]
I(r,b_{rel},b_{cm},R_0)
\nn
I(r,b_{rel},b_{cm},R_0)&\equiv&\int d{\bf r}'
\exp\left[-\frac{r'^2-2rr'\cos\theta}{b^2_{rel}}-\frac{r'^2+2rr'-4r'R_0}{(2b_{cm})^2}\right]\rho({\bf r}').
\nn&=&
2\pi\int_0^{\infty} r'^2dr'
\exp\left[-\frac{r'^2}{b^2_{rel}}-\frac{r'^2+2rr'-4r'R_0}{(2b_{cm})^2}\right]
\int_{-1}^1 d\cos\theta \exp\left[\frac{2rr'\cos\theta}{b^2_{rel}}\right]\rho(r',\cos\theta).
\nn&\approx&
\frac{b^2_{rel}}{4r}
\int_0^{\infty}r'dr'
\exp\left[-\frac{r'^2}{b^2_{rel}}-\frac{r'^2+2rr'-4r'R_0}{(2b_{cm})^2}\right]
\nn&\times&
\left[\exp{\left(\frac{2rr'}{b^2_{rel}}\right)}-
\exp{\left(-\frac{2rr'}{b^2_{rel}}\right)}\right]
\sum_a(2j_a+1)V^2_a{\cal R}^2_a(r')
\equiv\sum_a (2j_a+1)I_a(r,b_{rel},b_{cm},R_0).
\nn
\eea
\end{widetext}
Let us stress on the fact that the above general mean field expression
has a Woods--Saxon plus a Gaussian shape centered around $R_0$ given by the
integral $I$.
By replacing the monopole density with its mean value
\bea\label{rhohevi}
\rho_0(r')=\sum_a(2j_a+1)V^2_a{\cal R}^2_a(r')\rightarrow \frac{N_{\tau}}{R_{\tau}}\Theta(R_{\tau}-r')
\nn
\eea
where $R_{\tau}$ is the equivalent radius of the constant density distribution,
one obtains the integral in terms of the erf function
\bea
I(r,b_{rel},b_{cm},R_0)&=&I^{(+)}(r,b_{rel},b_{cm},R_0)\nn
&-&I^{(-)}(r,b_{rel},b_{cm},R_0)
\nn
\eea
where
\begin{widetext}
\bea
&&I^{(\pm)}(r,b_{rel},b_{cm},R_0)\equiv
\frac{N_{\tau}}{R_{\tau}}\frac{b^2_{rel}}{4r}\int_0^{R_{\tau}}r'dr'
\exp\left[-\frac{r'^2}{b^2_{rel}}-\frac{r'^2+2rr'-4r'R_0}{(2b_{cm})^2}
\pm\frac{2rr'}{b^2_{rel}}\right]
\nn&=&
\frac{N_{\tau}}{R_{\tau}}\frac{b^{2}_{rel}}{4r}
\left\{\frac{\sqrt{\pi}b^{\left(\pm\right)}}{4a^{\frac{3}{2}}}\exp\left(\frac{ \left(b^{\left(\pm\right)}\right)^{2}}{4a}\right)
\left[\textrm{erf}\left(\frac{2aR_{\tau}-b^{\left(\pm\right)}}{2\sqrt{a}}\right)+\textrm{erf}\left(\frac{b^{\left(\pm\right)}}{2\sqrt{a}}\right)\right]-
\frac{1}{2a}\left[\exp\left(R_{\tau}\left(b^{\left(\pm\right)}-aR_{\tau}\right)\right)-1\right]\right\}
\nn
\eea
\end{widetext}
with
\bea
a&=&\frac{1}{b^{2}_{rel}}+\frac{1}{\left(2b_{cm}\right)^{2}}
\nn
b^{\left(\pm\right)}&=&\pm\frac{2r}{b^{2}_{rel}}+\frac{4R_{0}-2r}{\left(2b_{cm}\right)^{2}}.
\eea
Using the obvious notation
\bea
I(r,b_{rel},b_{cm},R_0)&\equiv&
I_0(r,b_{rel},b_{cm},R_0)
\nn&+&
I_{cl}(r,b_{rel},b_{cm},R_0),
\eea
where the first term contains erf functions and the second one exponentials,
we can express the potential (\ref{VMF}) as follows
\bea
V_{MF}(r,b_{rel},b_{cm},R_0)&=&V_{0}(r)+V_{cl}(r)
\eea
where
\bea{\label{vzvcl}}
V_{0}(r)&=&
-v_0\exp\left[-\left(\frac{r}{b_{rel}}\right)^2-\left(\frac{r-2R_0}{2b_{cm}}\right)^2\right]
\nn&\times&
I_0(r,b_{rel},b_{cm},R_0)
\nn
V_{cl}(r)&=&
-v_0\exp\left[-\left(\frac{r}{b_{rel}}\right)^2-\left(\frac{r-2R_0}{2b_{cm}}\right)^2\right]
\nn&\times&
I_{cl}(r,b_{rel},b_{cm},R_0)
\nn&\equiv&
A_{cl}^{(-)}\exp\left[-\left(\frac{r-R_{cl}^{(-)}}{b_{cl}}\right)^2\right]
\nn&-&
A_{cl}^{(+)}\exp\left[-\left(\frac{r-R_{cl}^{(+)}}{b_{cl}}\right)^2\right]
\eea
with
\bea
A_{cl}^{(\pm)}&=&-v_{0}\frac{N_{\tau}}{R_{\tau}}\frac{b^{2}_{rel}}{2a}\frac{1}{4r}
\exp{\left[\frac{{R_{cl}^{(\pm)}}^{2}-R_{\tau}^{2}}{b_{cl}^{2}}-\frac{R_{0}^{2}-R_{0}R_{\tau}}{b_{cm}^{2}}
\right]}
\nn
R_{cl}^{(-)}&=&\frac{2R_{0}b^{2}_{rel}}{(2b_{cm})^{2}+b^{2}_{rel}}-R_{\tau}
=R_0\left(\frac{2}{y+1}-r_{\tau}\right)
\nn
R_{cl}^{(+)}&=&\frac{2R_{0}b^{2}_{rel}}{(2b_{cm})^{2}+b^{2}_{rel}}+R_{\tau}\left[\frac{(2b_{cm})^{2}-b^{2}_{rel}}{2b_{cm})^{2}+b^{2}_{rel}}\right]
\nn&=&
R_0\left(\frac{2}{y+1}+r_{\tau}\frac{y-1}{y+1}\right)
\nn
b_{cl}^{2}&=&\frac{(2b_{cm})^{2}b^{2}_{rel}}{(2b_{cm})^{2}+b^{2}_{rel}}
=\frac{(2b_{cm})^{2}}{y+1}
\nn
y&\equiv&\frac{(2b_{cm})^2}{b_{rel}^2}~,~
r_{\tau}\equiv\frac{R_{\tau}}{R_0}=0.75.
\eea
We used the systematic rules
\bea\label{rtrz}
R_{\tau}&=&1.2A_D^{1/3}
\nn
R_0&=&1.6A^{1/3}.
\eea
$R_{\tau}$ is the equivalent radius of the constant density distribution used
in the approximation of the density found in Eq. (\ref{rhohevi}). The simple 
parameterization given here, equivalent to the usual spherical nuclear saturation radius, 
was found to be valid for all the emitters studied in this work. $R_{0}$ is
parametrized here in terms of $A$, the mass number of the parent nucleus. The resulting
value is slightly beyond that of the geometrical touching radius of Eq. (\ref{geotr}) and is equivalent
with the parameterization given in the main text at the end of section III.1. 
The value was chosen for its universal validity across the calculations performed
in this work. 

A special case is given by $y=1$, i.e. $2b_{cm}=b_{rel}$, leading to the following values
\bea{\label{rclsp}}
R^{(-)}_{cl}&=&R_0(1-r_{\tau})<<R_0
\nn
R^{(+)}_{cl}&=&R_0
\nn
b_{cl}^2&=&2b_{cm}^2.
\eea
At $r=R_0$, these give
\bea
A^{(+)}_{cl}\approx-v_{0}\frac{N_{\tau}}{R_{\tau}}\frac{b^{4}_{rel}}{16R_0}
\exp\left[-\frac{(R_0-R_{\tau})^2}{2b_{cm}^2}\right].
\eea
We can rewrite the direct part of the mean field (\ref{VMF}) as the following summation
\bea
V_{MF}(r,b_{rel},b_{cm},R_0)&=&\sum_a(2j_a+1)
\nn&\times&V^{(a)}_{MF}(r,b_{rel},b_{cm},R_0),
\eea
in terms of the general function
\begin{widetext}
\bea
\label{VMFa}
V^{(a)}_{MF}(r,b_{rel},b_{cm},R_0)\equiv
-v_0\exp\left[-\left(\frac{r}{b_{rel}}\right)^2-\left(\frac{r-2R_0}{2b_{cm}}\right)^2\right]I_a(r,b_{rel},b_{cm},R_0).
\eea
\end{widetext}
Concerning the exchange part
one obtains for the first monopole leading term the following expresion
\bea
&&\int d{\bf r}'\Gamma^{(exc)}({\bf r,r}')\psi_{am}({\bf r}')
\nn&=&
-\int d{\bf r}'v({\bf r,r}')\rho({\bf r,r}')\psi_{am}({\bf r}')
\nn&=&
-\int d{\bf r}'v({\bf r,r}')\sum_bV_b^2
\sum_{\mu=-j_b}^{j_b}\psi_{b\mu}({\bf r}')\psi_{b\mu}({\bf r})\psi_{am}({\bf r}')
\nn&=&
-\int d{\bf r}'v({\bf r,r}')\sum_bV_b^2
\nn&\times&
\sum_{\mu=-j_b}^{j_b}{\cal R}_b(r'){\cal Y}^{\dag}_{j_b\mu}(\what{r'})
{\cal R}_b(r){\cal Y}_{j_b\mu}(\what{r}){\cal R}_a(r'){\cal Y}_{j_am}(\what{r'})
\nn&\approx&
-\int d{\bf r}'v({\bf r,r}')V_a^2\frac{1}{4\pi}
{\cal R}^2_a(r'){\cal R}_a(r){\cal Y}_{j_am}(\what{r})
\nn&=&
V_{MF}^{(a)}(r,b_{rel},b_{cm},R_0)\psi_a({\bf r}),
\eea
where we notice a smaller contribution given by only one $a$-th direct mean field term (\ref{VMFa}).

\end{document}